\documentclass[10pt,leqno]{amsart}
\usepackage[a4paper,margin=1in]{geometry}
\usepackage{graphicx} 
\usepackage{url}
\usepackage{multirow}%
\usepackage{array}%
\usepackage{amsmath,cite,url}
\usepackage{booktabs}%
\usepackage{breqn}
\usepackage[table]{xcolor}%
\usepackage{dblfloatfix}
\newcolumntype{P}[1]{>{\centering\arraybackslash}p{#1}}
\usepackage{hhline}
\usepackage{accents}%
\usepackage{algorithm}%
\usepackage{algorithmicx}%
\usepackage{algpseudocode}%
\usepackage{listings}%
\usepackage{caption}%
\usepackage{subcaption}%
\usepackage[T1]{fontenc}
\usepackage[utf8]{inputenc}
\usepackage{amsmath,amssymb,amsfonts}%
\usepackage{amsthm}%
\usepackage{mathrsfs}%
\usepackage{bm}
\theoremstyle{definition}

\theoremstyle{remark} 
 
\numberwithin{equation}{section}

\begin{document} 
\title[short title]{Graph Representation of RaagBase: A Unique Dataset for Hindustani Music} 
\author{Chandan Misra} 
\address[Chandan Misra]{XIM University, School of Computer Science and Engineering, 752050, 
Bhubaneswar, India} 
\email[Chandan Misra]{chandan@xim.edu.in} 
\author{Swarup Chattopadhyay} 
\address[Swarup Chattopadhyay]{XIM University, School of Computer Science and Engineering, 752050, 
Bhubaneswar, India} 
\email[Swarup Chattopadhyay]{swarupc@xim.edu.in} 
\subjclass[2000]{Primary xxx, yyyy; Secondary xxxx, yyyy} 
\date{June 1, 2017, accepted December 7, 2017.} 
\keywords{Dxxx gxxx, Axxx gxxx} 
\begin{abstract} 
Raag classification is a fundamental MIR task for Hindustani Music, with applications in recommendation, education, archiving, and intelligent search. However, raag clustering remains underexplored, as most existing approaches rely on annotated audio or labeled datasets. While annotated melodic phrases capture characteristic patterns, complete note sequences preserve temporal structure and contextual dependencies, making them more suitable for data-driven modeling. In this work, we introduce RaagBase, a notation-based text dataset consisting of note sequences from compositions by Pt. Bhatkhande. Furthermore we propose a novel graph-based representation of raag structures by modeling the dominance and absence of notes in compositions. Each composition is represented as a node, and  the edges between two compositions corresponds the similarities between them based on the note frequency distribution. Further, we apply established graph clustering techniques to identify groups of similar raag compositions. Experimental results demonstrate highly coherent clusters with strong agreement to ground-truth raag labels, thereby validating both the dataset and the proposed representation. The dataset is publicly available at \url{https://anonymous.4open.science/r/RaagBase-5427}.
\end{abstract} 
\maketitle 
\section{Introduction}
\label{sec:intro}

\textit{Hindustani Music}, also known as North Indian Classical Music, is one of the two major art music traditions of the Indian subcontinent. It is characterized by two primary frameworks: \textit{Raag}, the melodic framework, and \textit{Taal}, the rhythmic framework. Raag can be viewed as lying between a scale and a tune \cite{ganguli2018distributional}, providing a structured grammar that defines characteristic melodic sequences known as \textit{Aroh} (ascending) and \textit{Avroh} (descending). These sequences serve as the building blocks for constructing melodies that evoke specific moods and also provide cues for identifying raags.

Identification of a raag in any composition finds applications and research in music learning \cite{pendyala2022towards}, music information retrieval (MIR) \cite{chithra2015music,kirthika2012review,sridhar2009raga,murthy2018content}, automatic classification of moods of compositions \cite{ujlambkar2012mood,nag2022application}, recommending music to listeners, creating music generation systems, etc. Existing approaches to this problem primarily rely on annotated audio or music datasets \cite{srinivasamurthy2021saraga,gulati2016indian,shankar2024saraga} with labeled data primarily to train supervised models \cite{chowdhuri2019phononet,shah2021raga,gulati2016time}. However, such datasets are predominantly offer audio-based features, and require substantial manual effort and expert intervention for their creation \cite{shah2021raga}. While these datasets include annotated melodic phrases that capture characteristic patterns of a raag, the lack of complete note sequences limits data-driven statistical analysis, such as note frequency distribution, n-gram modeling \cite{ross2017identifying}, Hidden Markov Model \cite{pandey2003tansen} and transition probability estimation \cite{bhattacharjee2011hindustani}.

In this work, we propose an alternative perspective by formulating raag identification as a graph clustering problem over musical compositions. Rather than directly assigning labels, each composition is represented through its note frequency distribution derived from notated music sheets. Although improvisation in Indian art music introduces considerable variability in note sequences through ornamentation, timing, and phrase expansion—even within the same raag and composition—these variations largely preserve the underlying tonal distribution of notes, making frequency-based representations robust to performance-specific differences. The intuition behind this framework is that compositions belonging to similar raags share comparable sets of permissible notes (i.e., aroh and avroh), leading to similar tonal distributions. Consequently, they are expected to exhibit strong similarity measures and naturally form coherent clusters. Each such distribution is represented as a node in a graph, while edges capture the relationships between compositions based on the presence and absence of notes, thereby reflecting the underlying raag structure. In our work, this relationship is quantified using cosine similarity and Euclidean distance measures.

To support this framework, we introduce RaagBase, a dataset of Hindustani music compositions derived from one of the most authentic sources, \textit{Hindusthani Sangeet Paddhati-Kramik Pustak Malika}, which comprises approximately $1900$ compositions with notations and raags belonging to North Indian Classical Music penned by \textit{Pt. Vishnu Narayan Bhatkhande} ($1860$–$1936$). In order to reach a greater number of music students and scholars, the first volume of \textit{Kramik Pustak Malika} has been translated to \textit{Hindi} language in $1953$ by prominent music scholar \textit{Dr. Laxminarayan Garg}. We have used the music sheet compositions from Kramik Pustak Malika and have prepared a preliminary dataset, RaagBase, consisting of $116$ samples, each of which corresponds to the note frequency distribution of a composition, and label each such composition with the corresponding raags given in the book itself.

We visualize the communities in the proposed note frequency distribution–based graph model by applying state-of-the-art graph clustering techniques. We empirically demonstrate the robustness of RaagBase using multiple clustering performance indices across varying thresholds of the similarity measures, observing scores exceeding 90\% over a broad range of thresholds. Furthermore, we analyze bridge nodes connecting different communities and investigate their musical significance in terms of characteristic melodic sequences, namely aroh and avroh.

\section{The Creation and Description of RaagBase}
\label{sec:dataset-creation-description}
Pt. Vishnu Narayan Bhatkhande is the pioneer for providing a comprehensive theoretical foundation of Hindustani Sangeet in a published form in his six-volume book series titled \textit{Hindustani Sangeet Paddhati, Kramik Pustak Malika} written in \textit{Marathi} language in $1920$. His book contains a comprehensive description of music symbols for realizing musical components including notes (Svar), time signatures (Lay), beats (Taal), ornaments (Alankar) etc. The dataset created in the current work has been taken from the Hindi translation of the second volume of the series. The second volume of the book series contains a total of $319$ compositions belonging to $11$ different raags (as given in Table \ref{tab:raag-distribution}). The present work takes these written compositions as a source of musical information to create the database for North Indian Classical Music to be used in computational musicological research.

\begin{table}[]
    \centering
    \small
    \begin{tabular}{|c|c|c|}
        \hline
        \textbf{Sl. No.} & \textbf{Name of the Raag} & \textbf{Number of Compositions} \\
        \hline
        $1$ & Bhairav & $42$ \\
        \hline
        $2$ & Todi & $39$ \\
        \hline
        $3$ & Poorvi & $35$ \\
        \hline
        $4$ & Khamaj & $30$ \\
        \hline
        $5$ & Bilaval & $29$ \\
        \hline
        $6$ & Bhairavi & $27$ \\
        \hline
        $7$ & Marwa & $25$ \\
        \hline
        $8$ & Kafi & $24$ \\
        \hline
        $9$ & Yaman & $24$ \\
        \hline
        $10$ & Asavri & $24$ \\
        \hline
        $11$ & Yaman Kalyan & $20$ \\
        \hline
    \end{tabular}
    \caption{Raag Distribution of the compositions of the second volume of \textit{Hindustani Sangeet Paddhati-Kramik Pustak Malika}}
    \label{tab:raag-distribution}
\end{table}

We have taken $116$ compositions of the three highest frequent raags (more than $30$ compositions), i.e., raag \textit{Bhairav or bh}  ($42$), \textit{Todi or td} ($39$), and \textit{Poorvi or pv} ($35$), respectively, from the entire collection of $319$ compositions for the creation of machine learning and graph models and perform our experimental analysis. Eventually, the entire collection of compositions from all six volumes will be preserved in RaagBase, thereby enriching the dataset by including almost $1900$ compositions covering $50$ odd raags.

The dataset is a single \textit{comma separated values} (CSV) file where the notes of each composition along with the raags are manually stored. Therefore, each sample of the dataset corresponds to a single composition, and the notes of each composition span over three octaves or \textit{Saptak}s, namely the middle or \textit{Madhya}, upper or \textit{Taar}, and lower octave or \textit{Mandra Saptak}. Each octave consists of $12$ notes, which generates $36$ probable notes ($\mathcal{N}$ = $\{ \underaccent{\dot}{S}$, $S$, $\dot{S}$, $\underaccent{\dot}{r}$, $r$, $\dot{r}$, $\underaccent{\dot}{R}$, $R$, $\dot{R}$, $\underaccent{\dot}{g}$, $g$, $\dot{g}$, $\underaccent{\dot}{G}$, $G$, $\dot{G}$, $\underaccent{\dot}{M}$, $M$, $\dot{M}$, $\underaccent{\dot}{m}$, $m$, $\dot{m}$, $\underaccent{\dot}{P}$, $P$, $\dot{P}$, $\underaccent{\dot}{d}$, $d$, $\dot{d}$, $\underaccent{\dot}{D}$, $D$, $\dot{D}$, $\underaccent{\dot}{n}$, $n$, $\dot{n}$, $\underaccent{\dot}{N}$, $N$, $\dot{N} \}$) for each composition. We have indexed each note with an integer starting from $1$ and ending at $36$ as given in Table \ref{tab:notes-indices}.

\begin{table*}[!ht]
     \centering
     \small
    \begin{tabular}{|c|c|c|c|c|c|c|}
        \hline
        \textbf{Note} & \begin{tabular}{@{}c@{}}Shadaj \\ ($\underaccent{\dot}{S}$, $S$, $\dot{S}$)\end{tabular} & \begin{tabular}{@{}c@{}}Komal \\ Rishabh \\ ($\underaccent{\dot}{r}$, $r$, $\dot{r}$)\end{tabular} & \begin{tabular}{@{}c@{}}Suddha \\ Rishabh \\ ($\underaccent{\dot}{R}$, $R$, $\dot{R}$)\end{tabular} & \begin{tabular}{@{}c@{}}Komal \\ Gandhar \\ ($\underaccent{\dot}{g}$, $g$, $\dot{g}$)\end{tabular} & \begin{tabular}{@{}c@{}}Suddha \\ Gandhar \\ ($\underaccent{\dot}{G}$, $G$, $\dot{G}$)\end{tabular} & \begin{tabular}{@{}c@{}}Madhyam \\ ($\underaccent{\dot}{M}$, $M$, $\dot{M}$)\end{tabular} \\
        \hline
        \textbf{Indices} & 1, 13, 25 & 2, 14, 26 & 3, 15, 27 & 4, 16, 28 & 5, 17, 29 & 6, 18, 30 \\
        \hline
        \textbf{Note} & \begin{tabular}{@{}c@{}}Tivr \\ Madhyam \\ ($\underaccent{\dot}{m}$, $m$, $\dot{m}$)\end{tabular} & \begin{tabular}{@{}c@{}}Pancham \\ ($\underaccent{\dot}{P}$, $P$, $\dot{P}$)\end{tabular} & \begin{tabular}{@{}c@{}}Komal \\ Dhaivat \\ ($\underaccent{\dot}{d}$, $d$, $\dot{d}$)\end{tabular} & \begin{tabular}{@{}c@{}}Suddha \\ Dhaivat \\ ($\underaccent{\dot}{D}$, $D$, $\dot{D}$)\end{tabular} & \begin{tabular}{@{}c@{}}Komal \\ Nishad \\ ($\underaccent{\dot}{n}$, $n$, $\dot{n}$)\end{tabular} & \begin{tabular}{@{}c@{}}Suddha \\ Nishad \\ ($\underaccent{\dot}{N}$, $N$, $\dot{N}$)\end{tabular} \\
        \hline
        \textbf{Indices} & 7, 19, 31 & 8, 20, 32 & 9, 21, 33 & 10, 22, 34 & 11, 23, 35 & 12, 24, 36 \\
        \hline
    \end{tabular}  
    \caption{Notes and indices of lower octave or \textit{Mandra Saptak} (indices 1 to 12), Middle octave or \textit{Madhya Saptak} (indices 13 to 24), and upper octave or \textit{Taar Saptak} (indices 25 to 36)}
    \label{tab:notes-indices}
\end{table*}

Once such mapping is established we generate the note table for each such $116$ compositions. To create the final dataset we generate the frequency distribution of each composition which serves as $36$ independent variables and obtain the raag of the composition for the dependent variable. Table~\ref{tab:mapping} and Table~\ref{tab:frequency} shows the overall process of mapping a composition to a frequency table.


\begin{table*}[!ht]
    \centering
    \small
    \begin{tabular}{|c|c|c|c|c|c|c|c|c|c|c|c|c|c|c|c|}
        \hline
        14 & 17 & 14 & 17 & 18 & 20 & 17 & 18 & 24 & 21 & 21 & 17 & 17 & 18 & 17 & 14 \\
        \hline
        17 & 18 & 20 & 17 & 18 & 24 & 20 & 0 & 14 & 14 & 25 & 29 & 29 & 17 & 17 & 18 \\
        \hline
        24 & 21 & 14 & 14 & 21 & 21 & 14 & 14 & 17 & 18 & 20 & 17 & 18 & 17 & 14 & 12 \\
        \hline
    \end{tabular}
    \caption{Mapping from actual note to its corresponding integer value}
    \label{tab:mapping}
\end{table*}

\begin{table*}[!ht]
    \centering
    \small
    \begin{tabular}{|c|c|c|c|c|c|c|c|c|c|c|c|c|}
        \hline
        \multicolumn{13}{|c|}{\textbf{Lower Octave or Mandra Saptak}} \\
        \hline
        \textbf{Note Index} & 1 & 2 & 3 & 4 & 5 & 6 & 7 & 8 & 9 & 10 & 11 & 12 \\
        \hline
        \textbf{Frequency} & 0 & 0 & 0 & 0 & 0 & 0 & 0 & 0 & 0 & 0 & 0 & 1 \\
        \hline
        \multicolumn{13}{|c|}{\textbf{Middle Octave or Madhya Saptak}} \\
        \hline
        \textbf{Note Index} & 13 & 14 & 15 & 16 & 17 & 18 & 19 & 20 & 21 & 22 & 23 & 24 \\
        \hline
        \textbf{Frequency} & 0 & 10 & 0 & 0 & 13 & 8 & 0 & 4 & 5 & 0 & 0 & 3 \\
        \hline
        \multicolumn{13}{|c|}{\textbf{Upper Octave or Taar Saptak}} \\
        \hline
        \textbf{Note Index} & 25 & 26 & 27 & 28 & 29 & 30 & 31 & 32 & 33 & 34 & 35 & 36 \\
        \hline
        \textbf{Frequency} & 1 & 0 & 0 & 0 & 2 & 0 & 0 & 0 & 0 & 0 & 0 & 0 \\
        \hline
        \multicolumn{13}{|c|}{\textbf{$12$ Note Indices and their respective cumulative frequencies}} \\
        \hline
        \textbf{Note Index} & \cellcolor{blue!25}1 & \cellcolor{blue!25}2 & 3 & 4 & \cellcolor{blue!25}5 & \cellcolor{blue!25}6 & 7 & \cellcolor{blue!25}8 & \cellcolor{blue!25}9 & 10 & 11 & \cellcolor{blue!25}12 \\
        \hline
        \textbf{Frequency} & 1 & 10 & 0 & 0 & 15 & 8 & 0 & 4 & 5 & 0 & 0 & 4 \\
        \hline
    \end{tabular}
    \caption{Frequency table of individual notes}
    \label{tab:frequency}
\end{table*}

Instead of taking the frequency distribution of $36$ notes (i.e. set $\mathcal{N}$), we merge three similar notes belonging to three different octaves to obtain the frequency distribution of $12$ notes (i.e. set $\mathcal{N}_{12}$, where $\mathcal{N}_{12} = \bigl\{S, r, R, g, G, M, m, P, d, D, n, N \bigr\} \subset \mathcal{N}$) only as shown in Table \ref{tab:frequency}. Since, the positions of the notes of the \textit{\={A}roh} and \textit{Avroh} of any particular composition in different octaves do not affect the r\={a}g of the composition, we map corresponding notes of three octaves and make a sum of frequencies of corresponding notes of three octaves as given in Table \ref{tab:frequency}. 

\begin{figure}[!ht]
     \centering
     \begin{subfigure}[b]{0.45\textwidth}
         \centering
         \includegraphics[width=\textwidth]{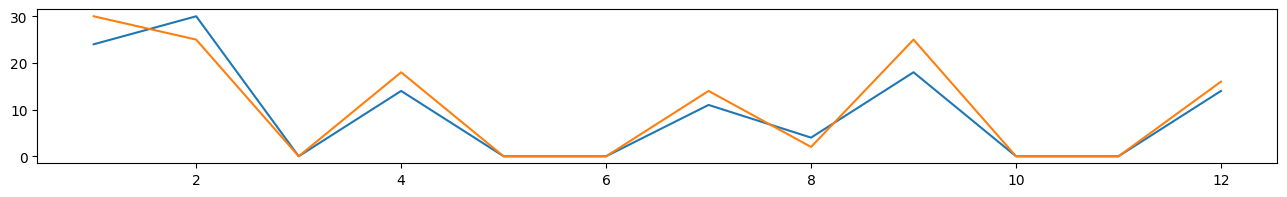}
         \caption{}
         \label{fig:same-raag-hist}
     \end{subfigure}
     \begin{subfigure}[b]{0.45\textwidth}
        \centering
        \includegraphics[width=\textwidth]{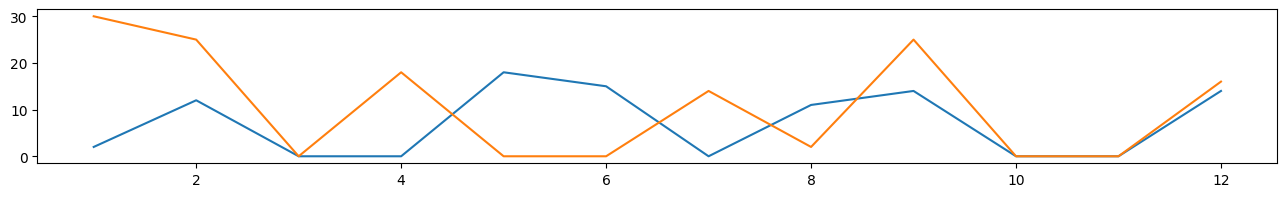}   
         \caption{}
         \label{fig:diff-raag-hist}
     \end{subfigure}
    \caption{Note frequency distribution across $12$ notes for \ref{fig:same-raag-hist}. two same raag (Todi)  and \ref{fig:diff-raag-hist}. two different raag (Bhairav and Todi)  compositions.}
    \label{fig:hist}
\end{figure}

Compositions belonging to same raag are the ones that follow the same Aroh and Avroh of the raag with occasional minute deviation. In other words the compositions must have a substantial note frequency on the note indices represented by the Aroh and Avroh of the raag the composition belongs to. 
Therefore, the note frequency distribution of the same raag compositions should follow the same silhouette. Each of Figure \ref{fig:same-raag-hist} and Figure \ref{fig:diff-raag-hist} overplots the note frequency distribution of a pair of compositions belonging to same (Bhairav) and different raag (Bhairav and Todi) respectively and shows similar and dissimilar silhouette of the note frequencies. To quantify the similarities between compositions having the same and different raag compositions, we have taken cosine similarity and Euclidean distance as a relationship measure. Since similar raag compositions are expected to exhibit the same similarity patterns, we expect higher values of cosine similarity and lower values of euclidean distance among compositions having the same raag. This analogy is experimentally validated and relationship scores are provided in Table \ref{tab:cos-ed}.

\begin{table}[!ht]
    \tiny
    \centering
    \begin{tabular}{|c|c|c|c|c|c|c|}
        \hline
        \textbf{$\frac{COS}{ED}$} & \textbf{$com_{bh}^{1}$} & \textbf{$com_{bh}^{2}$} & \textbf{$com_{pv}^{1}$} & \textbf{$com_{pv}^{2}$} & \textbf{$com_{td}^{1}$} & \textbf{$com_{td}^{2}$} \\
        \hline
        \textbf{$com_{bh}^{1}$} & $\frac{1.0}{0.0}$ & $\frac{0.888}{0.163}$ & $\frac{0.768}{0.237}$ & $\frac{0.701}{0.269}$ & $\frac{0.524}{0.341}$ & $\frac{0.462}{0.358}$ \\
        \hline
        \textbf{$com_{bh}^{2}$} & $\frac{0.888}{0.163}$ & $\frac{1.0}{0.0}$ & $\frac{0.771}{0.232}$ & $\frac{0.754}{0.239}$ & $\frac{0.628}{0.295}$ & $\frac{0.566}{0.315}$ \\
        \hline
        \textbf{$com_{pv}^{1}$} & $\frac{0.768}{0.237}$ & $\frac{0.771}{0.232}$ & $\frac{1.0}{0.0}$ & $\frac{0.971}{0.084}$ & $\frac{0.718}{0.261}$ & $\frac{0.687}{0.271}$ \\
        \hline
        \textbf{$com_{pv}^{2}$} & $\frac{0.701}{0.269}$ & $\frac{0.754}{0.239}$ & $\frac{0.971}{0.084}$ & $\frac{1.0}{0.0}$ & $\frac{0.706}{0.265}$ & $\frac{0.696}{0.266}$ \\
        \hline
        \textbf{$com_{td}^{1}$} & $\frac{0.524}{0.341}$ & $\frac{0.628}{0.295}$ & $\frac{0.718}{0.261}$ & $\frac{0.706}{0.265}$ & $\frac{1.0}{0.0}$ & $\frac{0.961}{0.097}$ \\
        \hline
        \textbf{$com_{td}^{2}$} & $\frac{0.462}{0.358}$ & $\frac{0.566}{0.315}$ & $\frac{0.687}{0.271}$ & $\frac{0.696}{0.266}$ & $\frac{0.961}{0.097}$ & $\frac{1.0}{0.0}$ \\
        \hline
    \end{tabular}
    \caption{Cosine Similarity (COS) and Euclidean Distance Dissimilarity (ED) relationship measures among same and different raag compositions.}
    \label{tab:cos-ed}
\end{table}

\section{Graph Representation of RaagBase}
\label{sec:graph-representation}
In this section we try to construct a graph from the compositions belonging to three different raags in RaagBase dataset. Each node $v^{r}_{i}$ of the graph represents a composition belonging to a particular raag, where $r \in \{bh, pv, td\}$ and $i \in \{0, 1, ..., 115\}$ i.e. $v^{bh}_{i}$, $i \in [0, 41]$ and $v^{pv}_{i}$, $i \in [42, 76]$ and $v^{td}_{i}$, $i \in [77, 115]$. Each edge of the graph provides a relationship measure between two nodes i.e. two compositions, represented as $E_{i,j}=rel(v^r_i, v^r_j)$. Therefore, an edge $E_{i,j}$ exists only when the relationship values is greater than some threshold i.e. $rel(v^r_i, v^r_j) > th$. Hence, different threshold values of the similarity measures produce varied dense graphs.

\begin{figure}[!ht]
     \centering
     \begin{subfigure}[b]{0.20\textwidth}
         \centering
         \includegraphics[width=\textwidth]{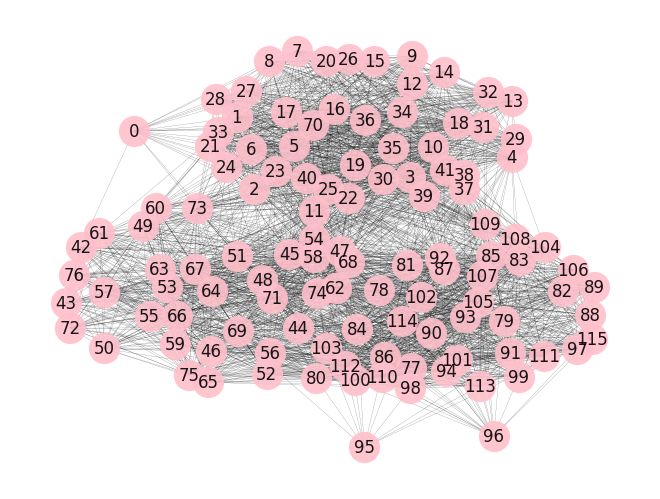}
         \caption{}
         \label{fig:graph-cos-8.png}
     \end{subfigure}
     \begin{subfigure}[b]{0.20\textwidth}
        \centering
        \includegraphics[width=\textwidth]{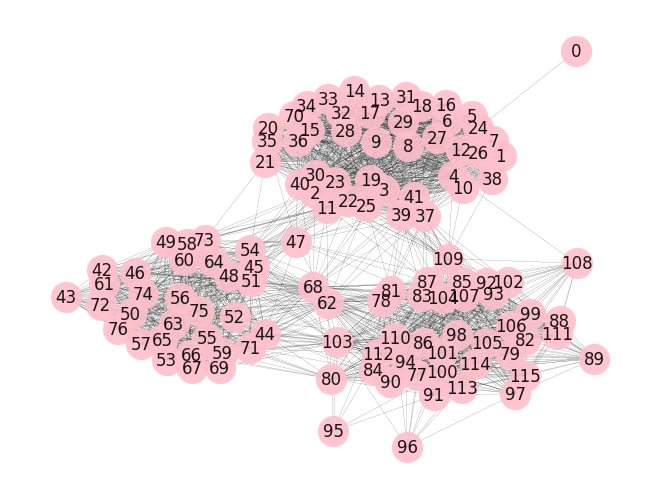}   
         \caption{}
         \label{fig:graph-cos-85}
     \end{subfigure}
     \begin{subfigure}[b]{0.24\textwidth}
        \centering
        \includegraphics[width=\textwidth]{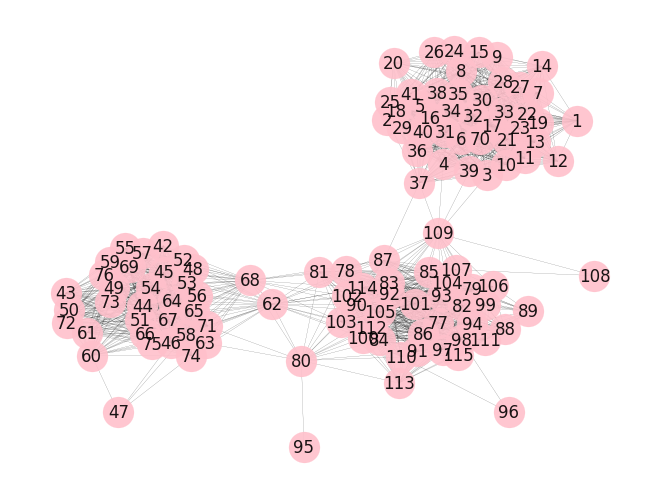}       
        \caption{}
        \label{fig:graph-cos-9}
     \end{subfigure}
     \begin{subfigure}[b]{0.20\textwidth}
        \centering
        \includegraphics[width=\textwidth]{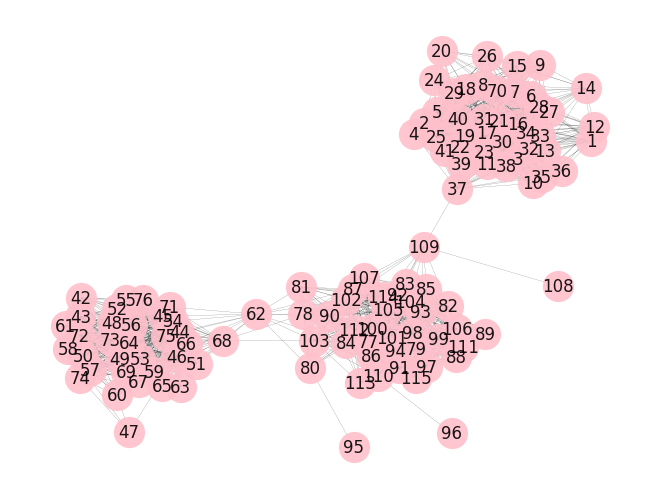}       
        \caption{}
        \label{fig:graph-cos-92}
     \end{subfigure}
     \begin{subfigure}[b]{0.20\textwidth}
         \centering
         \includegraphics[width=\textwidth]{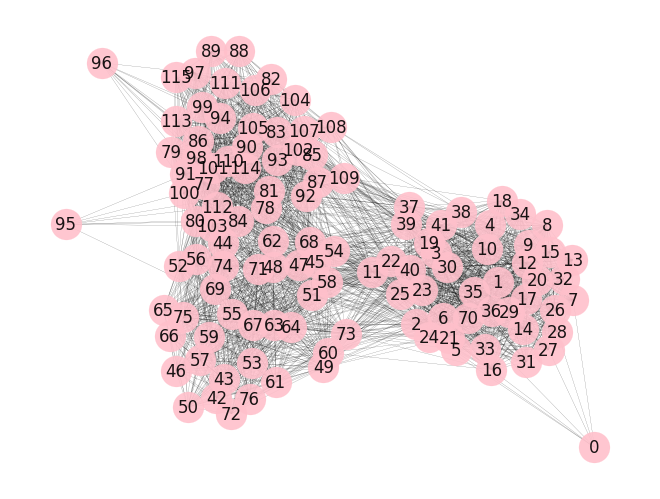}
         \caption{}
         \label{fig:graph-ed-0.24}
     \end{subfigure}
     \begin{subfigure}[b]{0.20\textwidth}
        \centering
        \includegraphics[width=\textwidth]{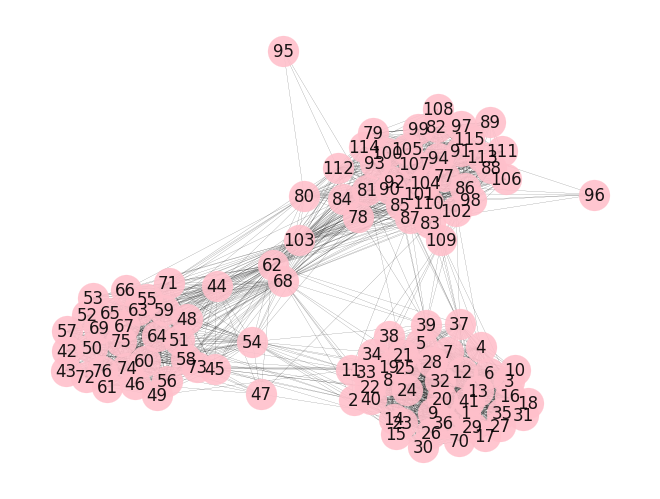}   
         \caption{}
         \label{fig:graph-ed-0.21}
     \end{subfigure}
     \begin{subfigure}[b]{0.20\textwidth}
        \centering
        \includegraphics[width=\textwidth]{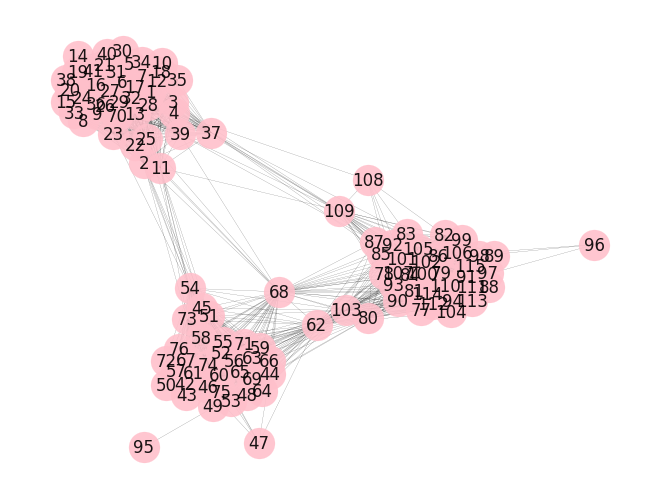}       
        \caption{}
        \label{fig:graph-ed-0.20}
     \end{subfigure}
     \begin{subfigure}[b]{0.20\textwidth}
        \centering
        \includegraphics[width=\textwidth]{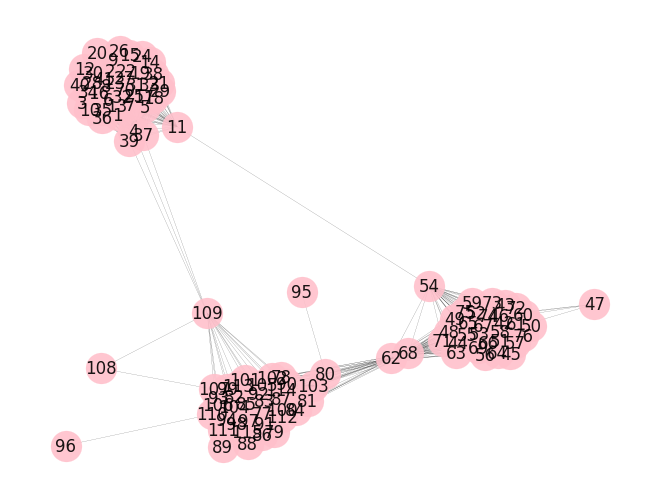}       
        \caption{}
        \label{fig:graph-ed-0.18}
     \end{subfigure}
    \caption{Generated graphs with increasing threshold values of the cosine similarity \ref{fig:graph-cos-8.png}. 0.80, \ref{fig:graph-cos-85}. 0.85, \ref{fig:graph-cos-9}. 0.90, and \ref{fig:graph-cos-92}. 0.92 and decreasing threshold values of Euclidean Distance \ref{fig:graph-ed-0.24}. 0.24, \ref{fig:graph-ed-0.21}. 0.21, \ref{fig:graph-ed-0.20}. 0.20, and \ref{fig:graph-ed-0.18}. 0.18}
    \label{fig:cos-ed-graph}
\end{figure}

We have used two relationship measures namely cosine similarity and Euclidean distance dissimilarity and construct the graphs representing RaagBase compositions with increasing and decreasing threshold relationship measure values respectively as shown in Figure \ref{fig:cos-ed-graph}. In section \ref{sec:experimental-evaluation}, we further apply various graph clustering algorithms on above mentioned graphs and try to analyze different communities obtained.

\section{Experimental Evaluation}
\label{sec:experimental-evaluation}
\begin{figure*}[!ht]
     \centering
     \begin{subfigure}[b]{0.30\textwidth}
         \centering
         \includegraphics[width=\textwidth]{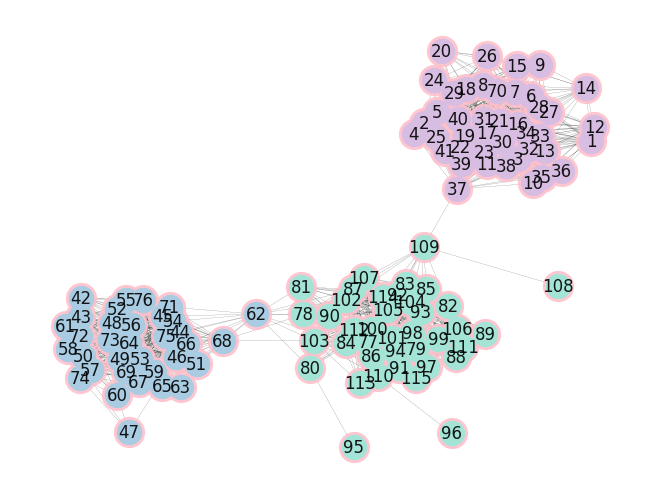}
         \caption{}
         \label{fig:lovain-cos-0.92}
     \end{subfigure}
     \begin{subfigure}[b]{0.30\textwidth}
        \centering
        \includegraphics[width=\textwidth]{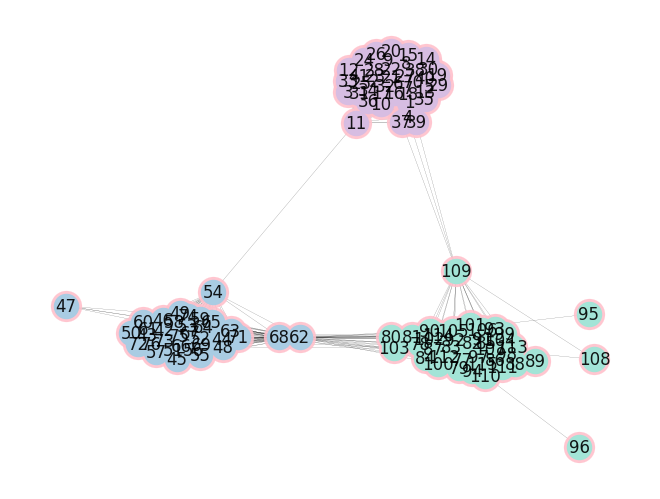}   
         \caption{}
         \label{fig:lovain-ed-0.18}
     \end{subfigure}
     \begin{subfigure}[b]{0.30\textwidth}
        \centering
        \includegraphics[width=\textwidth]{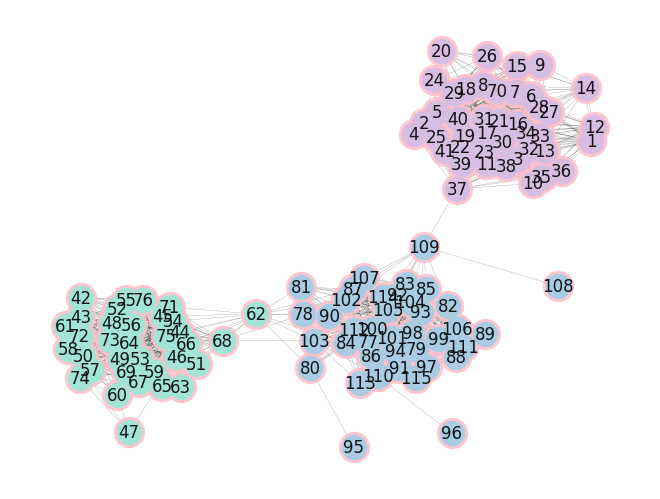}       
        \caption{}
        \label{fig:lpa-cos-0.92}
     \end{subfigure}
     \begin{subfigure}[b]{0.30\textwidth}
        \centering
        \includegraphics[width=\textwidth]{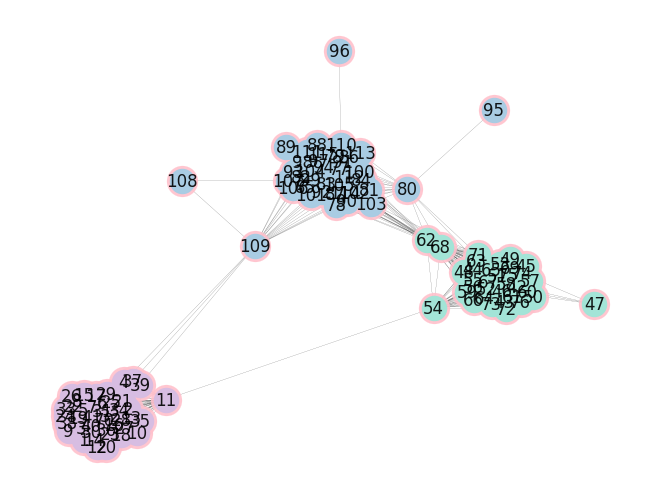}       
        \caption{}
        \label{fig:lpa-ed-0.18}
     \end{subfigure}
    \caption{Communities obtained using LA (\ref{fig:lovain-cos-0.92} and \ref{fig:lovain-ed-0.18}) and LPA algorithm (\ref{fig:lpa-cos-0.92} and \ref{fig:lpa-ed-0.18}) corresponding to a threshold values of $0.92$ and $0.18$ of cosine similarity and Euclidean Distance relationship measure respectively.}
    \label{fig:cos-ed-cluster-plot}
\end{figure*}

In this section we validate the graph representation of RaagBase by identifying similar raag compositions using two existing graph clustering algorithms. The first one is Louvain Algorithm (LA) \cite{blondel2008fast} based on modularity optimization technique and the other one is Label Propagation Algorithm (LPA) \cite{raghavan2007near} based on Random Walk. In order to measure the performance of the clustering algorithms, we have used three performance measures namely Normalized Mutual Information (NMI) \cite{strehl2002cluster}, Hubert-Arabie adjusted Rand Index (ARI) \cite{hubert1985comparing}, and Modularity (Mod) \cite{newman2006modularity} as given in Table \ref{tab:mop-cos} and \ref{tab:mop-eud}. We set the value of resolution parameter to 0.2 for computing Modularity. Additionally, we visualize the communities obtained by taking a pair of graph snapshots generated from applying LA and LPA with a higher and lower threshold values for cosine similarity and euclidean distance dissimilarity measure respectively as provided in Figure \ref{fig:cos-ed-cluster-plot}. As shown in Table \ref{tab:mop-cos}, the optimal NMI, ARI, and Mod values are attained at a cosine similarity threshold of 0.9, while preserving nearly the entire node set in the resulting network. This suggests the emergence of highly coherent raag clusters, corresponding to structurally significant communities. Consistent behavior is observed for Euclidean distance thresholds in the range of 0.15–0.20 (Table \ref{tab:mop-eud}), reinforcing the stability and robustness of the clustering across different similarity metrics. 

\begin{table}[!ht]
    \centering
    \tiny
    \begin{tabular}{|c|c|c|c|c|c|c|}
       \hline
       \multirow{2}*{\textbf{Threshold}} & \multirow{2}*{\textbf{\# Nodes}} & \multirow{2}*{\textbf{\# Edges}} & \multirow{2}*{\textbf{C.D.A.}} & \multicolumn{3}{|c|}{\textbf{M.P.}} \\
       \cline{5-7}
       & & & & NMI & ARI & Mod \\
       \hline
       \multirow{2}*{0.70} & \multirow{2}*{116} & \multirow{2}*{5241} & Louvain & 0.6572  & 0.5654 & 0.3992 \\
       \cline{4-7}
       & & & LPA & 0.0 & 0.0 & 0.7999 \\
       \hline
       \multirow{2}*{0.80} & \multirow{2}*{116} & \multirow{2}*{3050} & Louvain & 0.9634 & 0.9742 & 0.6553 \\
       \cline{4-7}
       & & & LPA & 0.0 & 0.0 & 0.7999 \\
       \hline
       \multirow{2}*{0.90} & \multirow{2}*{115} & \multirow{2}*{1819} & Louvain & 0.9632 & 0.9741 & 0.9156 \\
       \cline{4-7}
       & & & LPA & 0.8447 & 0.9872 & 0.9156 \\
       \hline
       \multirow{2}*{0.92} & \multirow{2}*{115} & \multirow{2}*{1600} & Louvain & 0.9632 & 0.9741 & 0.9245 \\
       \cline{4-7}
       & & & LPA & 0.8447 & 0.8567 & 0.9245 \\
       \hline
       \multirow{2}*{0.94} & \multirow{2}*{113} & \multirow{2}*{1269} & Louvain & 0.9625 & 0.9732 & 0.9277 \\
       \cline{4-7}
       & & & LPA & 0.9256 & 0.9482 & 0.9277 \\
       \hline
       \multirow{2}*{0.96} & \multirow{2}*{110} & \multirow{2}*{820} & Louvain & 0.9618 & 0.9724 & 0.9295 \\
       \cline{4-7}
       & & & LPA & 0.9618 & 0.9724 & 0.9295 \\
       \hline
       \multirow{2}*{0.98} & \multirow{2}*{91} & \multirow{2}*{251} & Louvain & 0.7294 & 0.5607 & 0.8142 \\
       \cline{4-7}
       & & & LPA & 0.7275 & 0.6361 & 0.8661 \\
       \hline
    \end{tabular}
    \caption{Measure of Performances (M.P.) of two community detection algorithms (C.D.A.) with increasing threshold values of cosine similarity.}
    \label{tab:mop-cos}
\end{table}

\begin{table}[!ht]
    \centering
    \tiny
    \begin{tabular}{|c|c|c|c|c|c|c|}
       \hline
       \multirow{2}*{\textbf{Threshold}}  & \multirow{2}*{\textbf{\# Nodes}} & \multirow{2}*{\textbf{\# Edges}} & \multirow{2}*{\textbf{C.D.A.}} & \multicolumn{3}{|c|}{\textbf{M.P.}} \\
       \cline{5-7}
       &  &  & & NMI & ARI & Mod \\
       \hline
       \multirow{2}*{0.40} & \multirow{2}*{116} & \multirow{2}*{6574} & Louvain & 0.5513 & 0.4922 & 0.4055\\
       \cline{4-7}
       & & & LPA & 0.0 & 0.0 & 0.7999 \\
       \hline
       \multirow{2}*{0.30} & \multirow{2}*{116} & \multirow{2}*{4943} & Louvain & 0.5474 & 0.4886 & 0.4794 \\
       \cline{4-7}
       & & & LPA & 0.0 & 0.0 & 0.8000 \\
       \hline
       \multirow{2}*{0.25} & \multirow{2}*{116} & \multirow{2}*{3066} & Louvain & 0.8151 & 0.8351 & 0.5301 \\
       \cline{4-7}
       & & & LPA & 0.7479 & 0.5999 & 0.6679 \\
       \hline
       \multirow{2}*{0.20} & \multirow{2}*{116} & \multirow{2}*{2017} & Louvain & 0.9632 & 0.9741 & 0.8653 \\
       \cline{4-7}
       & & & LPA & 0.9632 & 0.9741 & 0.8653 \\
       \hline
       \multirow{2}*{0.15} & \multirow{2}*{113} & \multirow{2}*{1468} & Louvain & 0.9256 & 0.9482 & 0.9240 \\
       \cline{4-7}
       & & & LPA & 0.9626 & 0.9732 & 0.9236 \\
       \hline
       \multirow{2}*{0.10} & \multirow{2}*{108} & \multirow{2}*{602} & Louvain & 0.9610 & 0.9714 & 0.9306 \\
       \cline{4-7}
       & & & LPA & 0.8771 & 0.8594 & 0.8711 \\
       \hline
       \multirow{2}*{0.08} & \multirow{2}*{91} & \multirow{2}*{267} & Louvain & 0.7924 & 0.7316 & 0.8937 \\
       \cline{4-7}
       & & & LPA & 0.7453 & 0.6750 & 0.8730 \\
       \hline
    \end{tabular}
    \caption{Measure of Performances (M.P.) of two community detection algorithms (C.D.A.) with decreasing threshold values of Euclidean Distance.}
    \label{tab:mop-eud}
\end{table}

It is clearly seen from Table that as we increase/decrease the threshold values of the cosine/euclidean relationship measure, the performance measure is also increasing. This suggests that with varying threshold values the algorithms are correctly identifying the actual clusters (with different node colors) present in a graph as obtained in Figure \ref{fig:cos-ed-cluster-plot}. The results therefore validates RaagBase as a robust music dataset that preserves the melodic properties of different raags in the compositions.

Additionally we want to attract our readers to few nodes in the graph which are correctly partitioned into the actual clusters. However, these nodes contain a number of edges with other nodes of a different cluster, thus creating a bridge between two clusters. For example, node $62$, $68$ are the bridge nodes between Poorvi and Todi raag clusters while node $109$ creates a bridge between Todi and Bhairav clusters. The existence of such bridge nodes can be explained in terms of the presence and/or absence of notes which are otherwise not permissible in the aroh and avroh of the raag concerned. As mentioned in section \ref{sec:dataset-creation-description} that a composition may not follow the Aroh and Avroh sequence of its raag very strictly and therefore may/may not contain notes in indices that not belong/belong to Aroh-Avroh. In other words, the composition may contain non-zero frequency on some indices not belong to Aroh-Avroh of the corresponding raag and vice-versa.

\begin{table*}[!ht]
    \centering
    \small
    \begin{tabular}{|c|c|c|c|c|c|c|c|c|c|c|c|c|}
        \hline
        \multicolumn{13}{|c|}{\textbf{Node Index 62}} \\
        \hline
        \textbf{Note Index} & \cellcolor{blue!25}1 & \cellcolor{blue!25}2 & 3 & 4 & \cellcolor{blue!25}5 & \cellcolor{blue!25}6 & \cellcolor{blue!25}7 & \cellcolor{blue!25}8 & \cellcolor{blue!25}9 & 10 & 11 & \cellcolor{blue!25}12 \\
        \hline
        \textbf{Frequency} & \cellcolor{blue!25}36 & \cellcolor{blue!25}14 & 0 & 0 & \cellcolor{blue!25}10 & \cellcolor{red!25}0 & \cellcolor{blue!25}27 & \cellcolor{blue!25}18 & \cellcolor{blue!25}29 & 0 & 0 & \cellcolor{blue!25}23 \\
        \hline
        \multicolumn{13}{|c|}{\textbf{Node Index 68}} \\
        \hline
        \textbf{Note Index} & \cellcolor{blue!25}1 & \cellcolor{blue!25}2 & 3 & 4 & \cellcolor{blue!25}5 & \cellcolor{blue!25}6 & \cellcolor{blue!25}7 & \cellcolor{blue!25}8 & \cellcolor{blue!25}9 & 10 & 11 & \cellcolor{blue!25}12 \\
        \hline
        \textbf{Frequency} & \cellcolor{blue!25}36 & \cellcolor{blue!25}19 & 0 & 0 & \cellcolor{blue!25}23 & \cellcolor{blue!25}4 & \cellcolor{blue!25}32 & \cellcolor{blue!25}26 & \cellcolor{blue!25}44 & \cellcolor{red!25}1 & 0 & \cellcolor{blue!25}41 \\
        \hline
        \multicolumn{13}{|c|}{\textbf{Node Index 109}} \\
        \hline
        \textbf{Note Index} & \cellcolor{blue!25}1 & \cellcolor{blue!25}2 & 3 & \cellcolor{blue!25}4 & 5 & 6 & \cellcolor{blue!25}7 & \cellcolor{blue!25}8 & \cellcolor{blue!25}9 & 10 & 11 & \cellcolor{blue!25}12 \\
        \hline
        \textbf{Frequency} & \cellcolor{blue!25}17 & \cellcolor{blue!25}8 & 0 & \cellcolor{blue!25}7 & 0 & \cellcolor{red!25}6 & \cellcolor{red!25}0 & \cellcolor{blue!25}4 & \cellcolor{blue!25}17 & 0 & 0 & \cellcolor{blue!25}11 \\
        \hline        
    \end{tabular}
    \caption{Frequency table for the bridge nodes $62$, $68$, and $109$. Blue cells indicate the Aroh-Avroh note indices and red cells indicates deviations from such indices.}
    \label{tab:improper-nodes}
\end{table*}

The existence or non-existence of notes in the Aroh-Avroh indices makes one raag composition closer to a different raag composition and creates inter-community edges. For example node $62$ contains zero frequency at note index $6$ which is a Aroh-Avroh note index (Suddha Madhyam) of raag Poorvi (shown in Table \ref{tab:improper-nodes}.) For the node $68$, we observe a non-zero note frequency at index $10$ which is a non Aroh-Avroh note index (Suddha Dhaibat) of raag Poorvi. We observe both types of improper note frequencies at note index $6$ and $7$ (Suddha Madhyam and Tivr Madhyam) in node $109$ belonging to raag Todi. These three notes produce inter-community bridges since they are impure compositions in the sense that they do not follow the permissible note sequences (Aroh and Avroh notes) in a strict way. We try to validate this assumption by manually converting these compositions to pure ones using introducing and/or removing improper notes from the frequency distribution.

\begin{figure}
    \centering
    \includegraphics[width=0.45\textwidth]{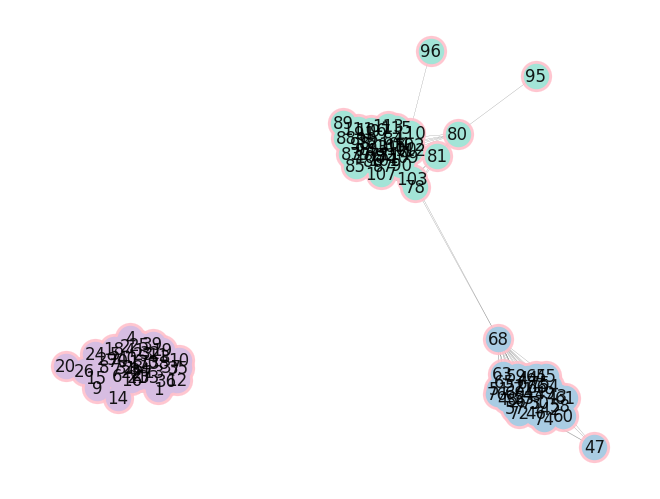}
    \caption{Clusters obtained after changing impure bridge compositions to pure compositions.}
    \label{fig:modified-clusters}
\end{figure}

We change the frequencies of node $62$ and $68$ by adding a positive frequency to note Suddha Madhyam and making the frequency of note Suddha Dhaivat zero respectively to convert them into pure Poorvi composition. Similarly, we add a positive frequency to note Tivr Madhyam and make the frequency of note Suddha Madhyam zero to convert node 109 to a pure Todi composition. We apply the same clustering algorithms to the modified graph (threshold value of cosine similarity of 0.92) and obtain the better raag clusters as shown in Figure \ref{fig:modified-clusters}.

It is clearly seen that we obtain better clusters after the modifications of the frequencies of the impure compositions with very less bridges, which can further be removed by making all the impure compositions to pure ones.

\subsection{Limitations:}
\label{sec:limitations}
We have discovered a couple of limitations in the our graph model. The first limitation is concerned about the instances where multiple raags share the same permissible notes, leaving uncertainty about the effectiveness of the proposed technique in such scenarios. For example, both Raag \textit{Bhoopali} and \textit{Deshkar} have the same notes in the aroh and avroh, it is easy to mistake one for the other. However, Deshkar is a member of the \textit{Bilawal} thaat family, whereas Bhoopali is a member of the \textit{Kalyan} thaat family. Their speed, mood, and note movement are all different. This kind of similarity in melodic pattern makes the raag identification difficult in string matching techniques as well and compositions may be differentiated based on other raag characteristics like \textit{Vadi}, \textit{Samvadi} etc. For example, raag Deshkar has Suddha Dhaivat and Suddha Gandhar as Vadi and Samvadi notes while for raag Bhoopali it is the opposite.

The second limitation is due to the inherent characteristics of similarity measures used. For example, cosine similarity would give same similarity score for two same and two different raag compositions if all the compositions have more or less similar frequency distributions. The reason is that cosine similarity is not sensitive to positions of the vector components. Therefore, all the three compositions would end up in the same cluster even if one of them is different from the other two.

\section{Conclusions and Future Work}
In this paper we created a music dataset RaagBase that consists of note frequency distributions of $116$ sample compositions belonging three different raags of North Indian Classical Music. We quantified the similarity between two same or different raag compositions using two relationship measures namely cosine similarity and Euclidean Distance Dissimilarity and showed that RaagBase exhibits intuitive outcomes i.e. high and low similarities and dissimilarities were found for two same raag compositions for cosine and euclidean respectively. The above result led us to create the graph representation of RaagBase for both the relationship measures and visually validate the fact that the same raag compositions gradually form groups in the graphs for varying threshold values of the relationship measures. We then applied standard graph clustering algorithms and observed that for a certain range of threshold values they give actual communities i.e. same raag compositions accurately. This range of threshold values may give valuable insight for classifying compositions in respective raags and other machine learning applications when the dataset contains all the Bhatkhande compositions. We are also interested to identify overlapping raag compositions for further musicological analysis after applying overlapping communities on the graph obtained from the optimal range of threshold values.

\bibliographystyle{ieeetr}
\bibliography{bibliography}
\end{document}